\documentclass{article}
\usepackage[utf8]{inputenc}
\usepackage{authblk}
\usepackage{setspace}
\usepackage[margin=1.25in]{geometry}
\usepackage{graphicx}
\graphicspath{ {./figures/} }
\usepackage{subcaption}
\usepackage{amsmath}
\usepackage{bm}
\usepackage{siunitx}
\usepackage{lineno}
\usepackage{hyperref}
\usepackage{upgreek}

\usepackage{float}
\usepackage{xcolor}
\usepackage{caption}
\usepackage[style=nature, 
citestyle=numeric-comp,
sorting=none]{biblatex}
\title{\textbf{Direct spatiotemporal imaging of carriers reveals a long-lived bulk photovoltaic mechanism}}

\author[1]{Saptam Ganguly}
\author[1,2]{Sebin Varghese}
\author[3]{Aaron M. Schankler}
\author[3]{Xianfei Xu}
\author[3]{Kazuki Morita}
\author[4]{Michel Viret}
\author[3]{Andrew M. Rappe}
\author[1,5,*]{Gustau Catalan}
\author[1,2,*]{Klaas-Jan Tielrooij}

\affil[1]{Catalan Institute of Nanoscience and Nanotechnology (ICN2), CSIC and BIST, Campus UAB, Bellaterra, Barcelona, 08193 Spain.}
\affil[2]{Department of Applied Physics, TU Eindhoven, Den Dolech 2, 5612 AZ Eindhoven, The Netherlands.}
\affil[3]{Department of Chemistry, University of Pennsylvania, Philadelphia, Pennsylvania 19104-6323, USA.}
\affil[4]{SPEC, CEA, CNRS, Universit\'{e} Paris-Saclay, Gif-sur-Yvette, France.}
\affil[5]{ICREA - Instituci\'o Catalana de Recerca i Estudis Avançats, Barcelona, Catalonia.}

\affil[*]{Correspondence to: gustau.catalan@icn2.cat, k.j.tielrooij@tue.nl}

\date{}

\begin{document}

\maketitle

\begin{abstract}
\textbf{The bulk photovoltaic effect (BPVE), a manifestation of broken centrosymmetry, has attracted interest as a probe of the symmetry and quantum geometry of materials, and for use in photovoltaic and optoelectronic devices. However, so far the effect has not been captured directly in space and time. Here, we use contactless pump-probe microscopy to visualize the spatiotemporal evolution of photoexcited carriers in single-crystal, mono-domain BiFeO\textsubscript{3}, a prototypical ferroelectric material. We observe asymmetric carrier transport along the polar axis, which confirms the intrinsic bulk- and symmetry-driven nature of the BPVE. Remarkably, this asymmetric transport persists for several nanoseconds after photoexcitation, which cannot be explained by conventional short-lived shift or phonon ballistic current BPVE mechanisms. Our Monte Carlo simulations show that asymmetric momentum scattering by defects, such as oxygen vacancies, leads to long-lived asymmetric carrier transport, as observed experimentally. Beyond fundamental insights, this paves the way towards controlling symmetry- and defect-driven photoresponses.}
\end{abstract}
\vspace{1cm}



The bulk photovoltaic effect (BPVE) generates a dc photocurrent under uniform illumination in homogeneous, unbiased materials lacking inversion symmetry, such as BiFeO\textsubscript{3} and BaTiO\textsubscript{3}. Therefore, these materials do not require a p-n junction or a built-in field, unlike in conventional photovoltaics \cite{sturman2021photovoltaic}. The generated photovoltage can exceed the limit set by the band gap of the material, which could allow the BPVE to overcome the Shockley-Queisser limit of photovoltaic efficiency \cite{spanier2016power, pusch2023energy}. As a result, the BPVE has attracted great attention for the development of next-generation optoelectronic and photovoltaic devices. The two prevailing microscopic explanations for the BPVE are the shift current mechanism and the ballistic current mechanism \cite{belinicher1980photogalvanic, belinicher1988relation, sturman1992}. 
\\

According to the shift current mechanism, the wavefunctions of photoexcited electrons experience an asymmetric redistribution in real space (the ``shift'') during the light-induced transition from the valence band to the conduction band. This asymmetric redistribution of the density matrix leads to a net current \cite{sipe2000second, dai2023recent}. This shift current is set by the Berry connection of Bloch states across the Brillouin zone, which quantifies the changing character of the Bloch states. Since this is allowed in materials with broken inversion symmetry, the shift current is inherently linked to the asymmetry of a crystal structure \cite{vonBaltz1981theory, Belinicher82p649, uzan2024observation} and has connections to wavefunction quantum geometry~\cite{Nagaosa17p1603345}. The shift current is thus a powerful tool for detecting polarity, quantum geometric effects \cite{akamatsu2021van, krishna2025terahertz, ahn2020low} and quantum phase transitions \cite{yang2024driving}; designing photovoltaic devices based on van der Waals materials and heterostructures \cite{liang2023strong,zeng2024dual}, topological insulators \cite{tan2016enhancement}, and semi-metals \cite{osterhoudt2019colossal, wang2019robust}; as well as observing phenomena, such as the circular \cite{de2017quantized, ji2019spatially} and surface photogalvanic effect \cite{xie2024surface}. 
\\

The ballistic current mechanism describes a net current that can be extracted due to an asymmetric carrier distribution in momentum space \cite{sturman2020ballistic, belinicher1988relation}: Along a broken symmetry axis, the probability of photo-generating carriers with forward momentum is different from the probability of generating carriers with the opposite momentum. The difference can arise from asymmetric scattering mechanisms such as electron-phonon or electron-electron interactions \cite{dai2023recent, dai2021phonon}. An additional imbalance in carrier populations -- leading to linear injection current -- can occur when the magnetic order breaks time-reversal symmetry. The key difference between shift currents and ballistic currents is that the former is due to an asymmetric \textit{real-space} redistribution of the electronic wavefunction upon excitation, while the latter is produced from an initially asymmetric \textit{momentum} distribution. Both shift and ballistic currents can coexist and contribute to the total photocurrent.
\\


The relevant time scales for both mechanisms lie in the sub-picosecond range \cite{gu2017mesoscopic}. For the shift current, the relevant time scale is the electron wavefunction decoherence time, while for ballistic current it is the momentum relaxation time. In the case of ferroelectrics, like BiFeO\textsubscript{3}, these time scales are below 10 fs and 1 ps, respectively \cite{gu2017mesoscopic, burger2019direct}. Such short lifetimes make the direct observation of photocarriers participating in the bulk photovoltaic effect highly challenging. Although the cumulative effect under constant illumination has been observed as a net BPVE voltage or current, no direct real-time and/or real-space observations of the photocarrier evolution giving rise to the BPVE have been reported. So far, only indirect signatures have been observed experimentally, such as terahertz emission in wurtzites \cite{laman2005ultrafast}. 
As a result, there is an important gap in our understanding of the microscopic mechanisms underlying the bulk photovoltaic effect. 
\\

\begin{figure}[H]
    \centering
    \includegraphics[width=\textwidth]{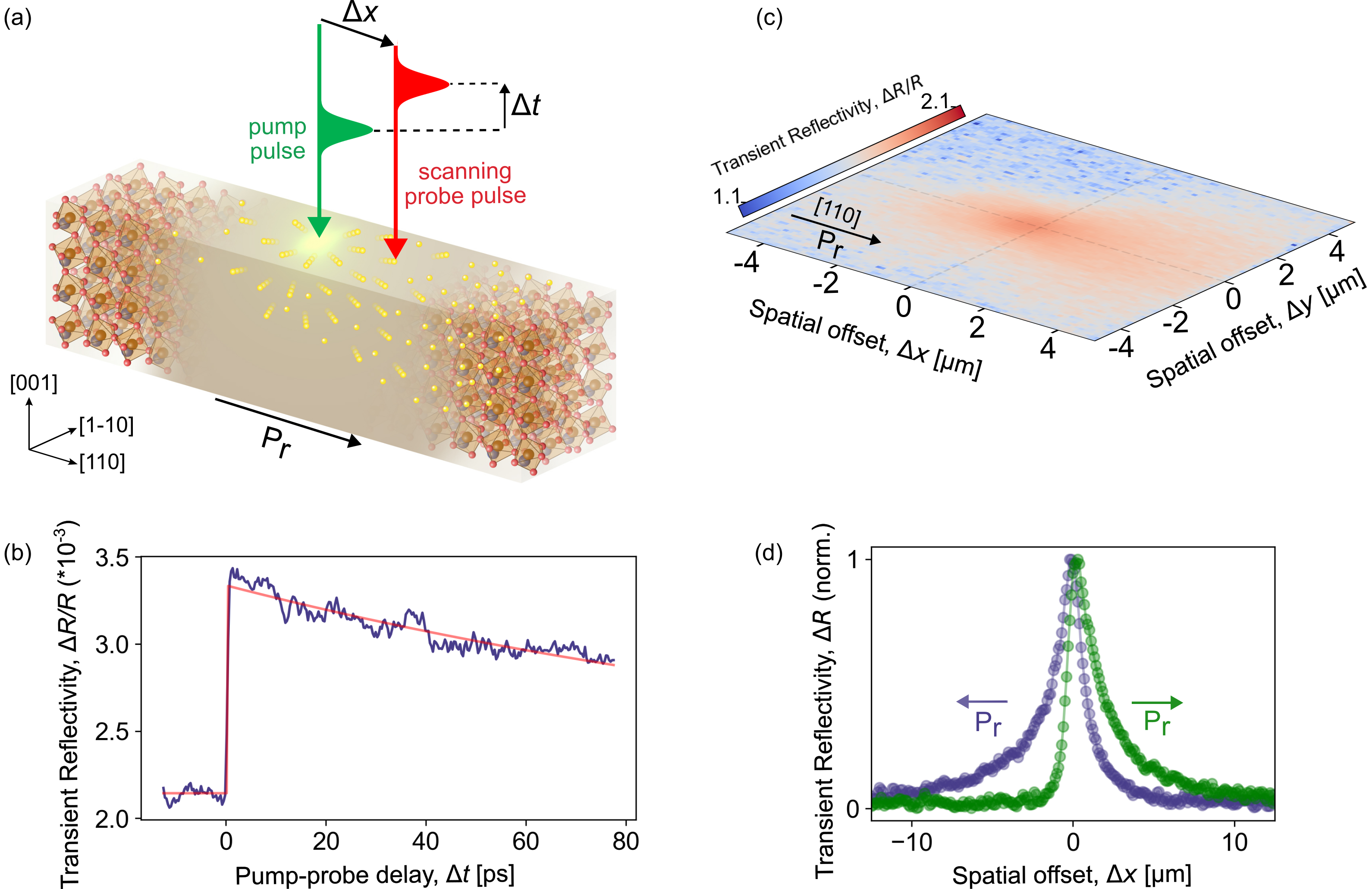}
    \caption{\textbf{Observing asymmetric charge transport with spatiotemporal microscopy.}
    \textbf{a)} Schematic representation of asymmetric transport of photoexcited charge carriers (yellow spheres) in BiFeO\textsubscript{3}. Spatiotemporal microscopy measurements allow for the  observation of this charge transport phenomenon directly in space and time. In this approach, a fixed pump pulse (green arrow) generates photoexcited charge carriers, while a probe pulse with controllable spatial offset and temporal delay (red arrow) detects photoexcited charges through a change in reflectivity $\Delta R$. \textbf{b)} Change in reflectivity of probe pulses $\Delta R$ as a function of pump-probe delay $\Delta t$, showing a rapid rise when the pump and  probe pulses overlap at $\Delta t$ = 0. We describe the data (blue line) with a \textit{Heaviside} step function convoluted with an exponential decay (red line), representing the ultrafast excitation of photoexcited charges, followed by their slow ($>$100 ps) recombination. 
    \textbf{c)} Pump-induced transient reflectivity map as a function of position at a pump-probe delay time of $\approx$13 ns. The pump is incident at the position $\Delta x = \Delta y$ = 0. This map represents the pump-induced change in charge carrier density, and the asymmetry indicates that photoexcited charge carriers spread further along the ferroelectric polarization direction $\vec P_r$ than in the opposite or perpendicular direction. 
    \textbf{d)} Spatial profiles of the transient reflectivity as a function of pump-probe offset $\Delta x$ in the case that the ferroelectric polarization points in the $-\Delta x$-direction (purple) and after rotating the crystal by 180$^\circ$, such that the ferroelectric polarization points in the $+\Delta x$-direction (green). Photoexcited charges consistently spread more in the direction of the ferroelectric polarization. 
}
    \label{fig:1}
\end{figure}

Here we address these challenges using contactless spatiotemporal pump–probe microscopy to visualize photoexcited carrier transport in a mono-domain single crystal of BiFeO\textsubscript{3}, and rationalize our observations with microscopic theory. Ultrafast spatiotemporal pump-probe microscopy enables the direct tracking of photogenerated charges in space and time with nanometer spatial resolution and femtosecond time resolution without any contacts \cite{Vazquez2024}, see Figs.~\ref{fig:1}a-b. The experimental results indicate charge transport that is clearly asymmetric along the polar crystal axis (see Figs.~\ref{fig:1} c-d), and that persists for many nanosecond after photoexcitation. This long duration of asymmetric transport is not compatible with the sub-picosecond timescale of shift and ballistic current mechanisms. Instead, we propose that persistent asymmetric charge spreading is maintained by scattering with inherently asymmetric defect states.
Using Monte Carlo calculations, we simulate asymmetric momentum scattering of carriers by defects, and demonstrate that this mechanism can provide sustained carrier asymmetry over the observed nanosecond duration, while also reproducing the experimentally obtained drift velocity $v_{\rm d}$ of $\approx \qty{50}{\m/\s}$. We therefore suggest that much of the macroscopically observed BPVE in BiFeO\textsubscript{3} is microscopically governed by asymmetric defect scattering. 
This defect-driven BPVE likely plays a role in other defect-prone polar materials, such as perovskite ferroelectrics, and opens up the possibility to control BPVE responses via defect engineering. 
\\

\section*{Results}
\subsection*{Spatiotemporal mapping of charge spreading in BiFeO\textsubscript{3} crystals}

For our experiments, we synthesized single crystals of mono-domain BiFeO\textsubscript{3} using the technique reported by Lebeugle \textit{et al.} \cite{lebeugle2007room}. BiFeO\textsubscript{3} has a rhombohedral crystal structure with the ferroelectric polarization along the pseudo-cubic [111] direction (see \ref{fig:ext1}a). The crystal was grown in the form of platelets, such that the pseudo-cubic [001] direction is perpendicular to the surface and the [110] direction is parallel to the crystal $x$-axis (see \ref{fig:ext1}b). We performed structural characterization of the sample using Raman, X-ray diffraction, and piezoresponse force microscopy to confirm the single orientation of the ferroelectric polarization in the single crystal (see \ref{fig:ext1}c-e). The single-crystal and single-domain character enables the unambiguous correlation of the optical response with the polar orientation of the crystal and avoids potentially confounding contributions from domain walls.
We also performed optoelectronic measurements (see Methods) which demonstrate that the crystal exhibits clear signatures of the BPVE, as shown in \ref{fig:ext2}. 
\\

To understand the microscopic origin of this BPVE response, we track the photoexcited charges in both time and space, using a home-built pump-probe microscopy setup (see Methods and \ref{fig:ST} for details). Pump pulses with a wavelength $\lambda_{\rm pump}$ of 515 nm and a pulse duration of $\approx$150 fs generate photoexcited charges nearly instantaneously. Equally short probe pulses, with a wavelength $\lambda_{\rm probe}$ of 770 nm, record how these photoexcited charges spread in space and in time. We achieve this by controlling the spatial offset between pump and probe pulses, $\Delta x$ and $\Delta y$, using a scanning mirror, and the temporal delay, $\Delta t$, using an optical delay line. The probe pulses detect photogenerated charges through the pump-induced change in the reflectivity of the sample, \textit{i.e.}\ the transient reflectivity $\Delta R$. This technique thus enables the direct observation of ultrafast transport of photogenerated carriers as they move away from where they were initially created, see Fig.\ \ref{fig:1}a. 
\\

We first perform pump-probe measurements where the pump and probe overlap in space ($\Delta x = \Delta y$ = 0), varying only the time delay $\Delta t$. The results show slow decay with a first relaxation time of at least 100 ps, see Fig.\ \ref{fig:1}b. The substantial signal before $\Delta t$ = 0, indicates that overall relaxation takes longer than the 13 ns delay between subsequent pump pulses in the pulse train. Additional time-resolved photoluminescence measurements (\ref{fig:ext3}a) confirm that photogenerated charges are long-lived, as the signal decays biexponentially with time constants of 4.2 and 20.2 ns.  We also verify that the transient reflectivity increases linearly with incident pump power, see \ref{fig:ext4}b, which means that the transient reflectivity profiles directly represent carrier density profiles and do not include nonlinear effects. 
\\

We now examine the spatial distribution of the photo-generated carriers. Figure \ref{fig:1}c shows a two-dimensional transient reflectivity map $\Delta R$ ($\Delta x, \Delta y$), where we scan the probe pulses along the in-plane $(\Delta x, \Delta y)$-directions. The map is clearly asymmetric along the $\Delta x$-direction, which corresponds to the direction of the in-plane component of the ferroelectric polarization. In contrast, the map is symmetric along the $\Delta y$-direction, which is perpendicular to the in-plane component of the polarization. To verify that the asymmetry along the polar axis is not an artifact coming from \textit{e.g.}\ a crystal tilt, we measure two spatial line scans along the x-axis, where we rotated the crystal by \ang{180}, see Figure \ref{fig:1}d. Both line scans show clear asymmetry, correlated with the macroscopic orientation of the crystal, thus confirming the directional dependence of photogenerated charge spreading, dictated by the polarization direction. The transient reflectivity map and the line-cuts in Fig.\ \ref{fig:1} correspond to a pump-probe time delay $\Delta t$ of 13 ns, which is more than three orders of magnitude longer than the sub-picosecond duration of either shift or ballistic photocurrent effects. This suggests the presence of a long-lived asymmetric transport mechanism.
\\

\subsection*{Quantifying asymmetric charge spreading}

\begin{figure}[H]
    \centering
    \includegraphics[width=\linewidth]{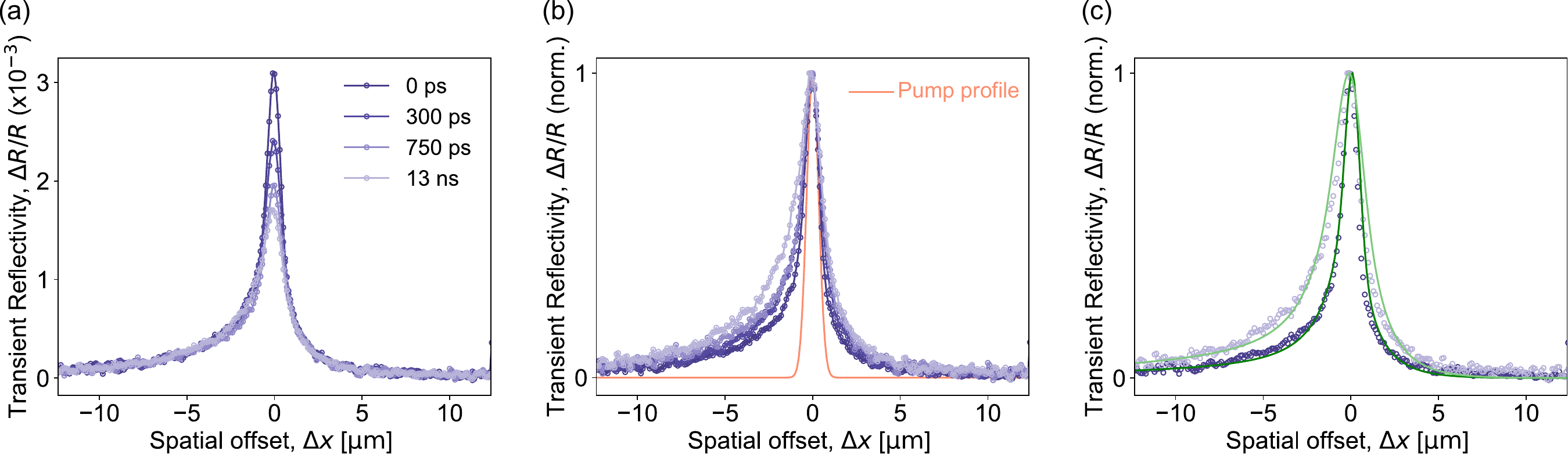}
    \caption{\textbf{Quantifying asymmetric charge spreading in space and time.}
    \textbf{a-b)} Spatial profiles (\textbf{a}) and normalized spatial profiles (\textbf{b}) of the transient reflectivity measured at different pump–probe delay times $\Delta t$ along the polar direction of BiFeO$_3$. The profiles exhibit a persistent asymmetry toward the $-x$ direction and a large background signal that indicates that photoexcited carriers exist longer than the time interval between subsequent pulses, which is 13 ns.
    \textbf{c)} Drift–diffusion model (dashed green lines) compared with experimental profiles at $\Delta t = 0$ ps and $\Delta t = 13$ ns. The simulation ($\alpha = 0.06$ cm$^2$/s, $v_d = 50$ m/s, $\tau = 500$ ns) reproduces both the symmetric broadening and the asymmetric displacement.}
    \label{fig:Data_Sim}
\end{figure}

To quantify how the photogenerated carriers spread in space, we measure transient reflectivity profiles 
$\frac{\Delta R}{R}(\Delta x)$ at different pump–probe delay times $\Delta t$, as shown in Fig.~\ref{fig:Data_Sim}a. For conventional diffusion, an initially Gaussian excitation profile remains Gaussian at later times and 
broadens symmetrically with increasing delay time~\cite{Vazquez2024,zheng2026photothermal}. However, for BiFeO$_3$ measured along the polar direction, the spatial profiles in Fig.~\ref{fig:Data_Sim}a remain asymmetric at all delay times. The signal extends more strongly toward the $-x$ direction than toward the $+x$ direction. In particular, we observe a large asymmetric background in the spatial profiles, which is present before $\Delta t$ = 0. Since the signal does not fully decay between pulses, a residual contribution remains when a next pump pulse arrives, corresponding to an effective delay time $\Delta t \approx$ 13 ns. To isolate spatial broadening from overall signal decay, we normalize the spatial profiles at each pump–probe delay time (Fig.~\ref{fig:Data_Sim}b). This normalization corrects for the temporal decay of the signal, yet does not remove the asymmetric background contribution. After normalization, the profiles clearly broaden with increasing delay time. Importantly, the broadening is not symmetric. This directional asymmetry suggests the presence of an effective asymmetric drift component superimposed onto symmetric spreading due to conventional diffusion. 
\\

To obtain a first estimate of the velocity of the drift component, we describe the spatial carrier density profiles using two Gaussian functions, where one has its centre fixed at $\Delta x$ = 0 and the other is not constrained (\ref{fig:ext6}). The centered Gaussian corresponds to photoexcited carriers that undergo symmetric diffusion, while the unconstrained Gaussian corresponds to carriers that undergo asymmetric transport. We compare the peak position of the latter component between $\Delta t$ = 0 and $\Delta t$ = 13 ns, finding a displacement of 600 nm. This suggests that the peak position of the asymmetric component between 0 ps and 13 ns moves at a speed of $\approx 50$ m/s. 
\\

In order to gain more insight into the underlying transport mechanism, we model the spatiotemporal evolution of the photoexcited carrier population considering a combination of drift, diffusion, and decay (see Methods: Phenomenological population dynamics model). Here, photoexcited charge carriers diffuse symmetrically and drift asymmetrically, while decaying by recombination or relaxation. We consider two populations of charge carriers, one that decays and diffuses symmetrically and one that drifts in addition to diffusion and decay. We compare the simulated profiles with the experimental data at $\Delta t = 0$ ps and $\Delta t = 13$ ns (Fig.~\ref{fig:Data_Sim}c) and obtain a consistent description of both symmetric broadening and asymmetric displacement when using a charge diffusivity of $\alpha = 0.06$ cm$^2$/s, a drift velocity of $v_d = 50$ m/s, and a decay time of $\tau = 500$ ns. The drift velocity is in good agreement with the velocity obtained using the description with two Gaussian populations (\ref{fig:ext6}). The obtained diffusivity value agrees with the scattering rates of other oxide perovskites~\cite{dai2021phonon}. The long lifetime reflects the presence of a slow decay component, consistent with the persistent background signal (see also Fig.~\ref{fig:1}b) and likely associated with defect-related states. 
\\





\subsection*{Simulations of defect-induced asymmetric charge spreading}

\begin{figure}[H]
    \centering
    \includegraphics[width=\textwidth]{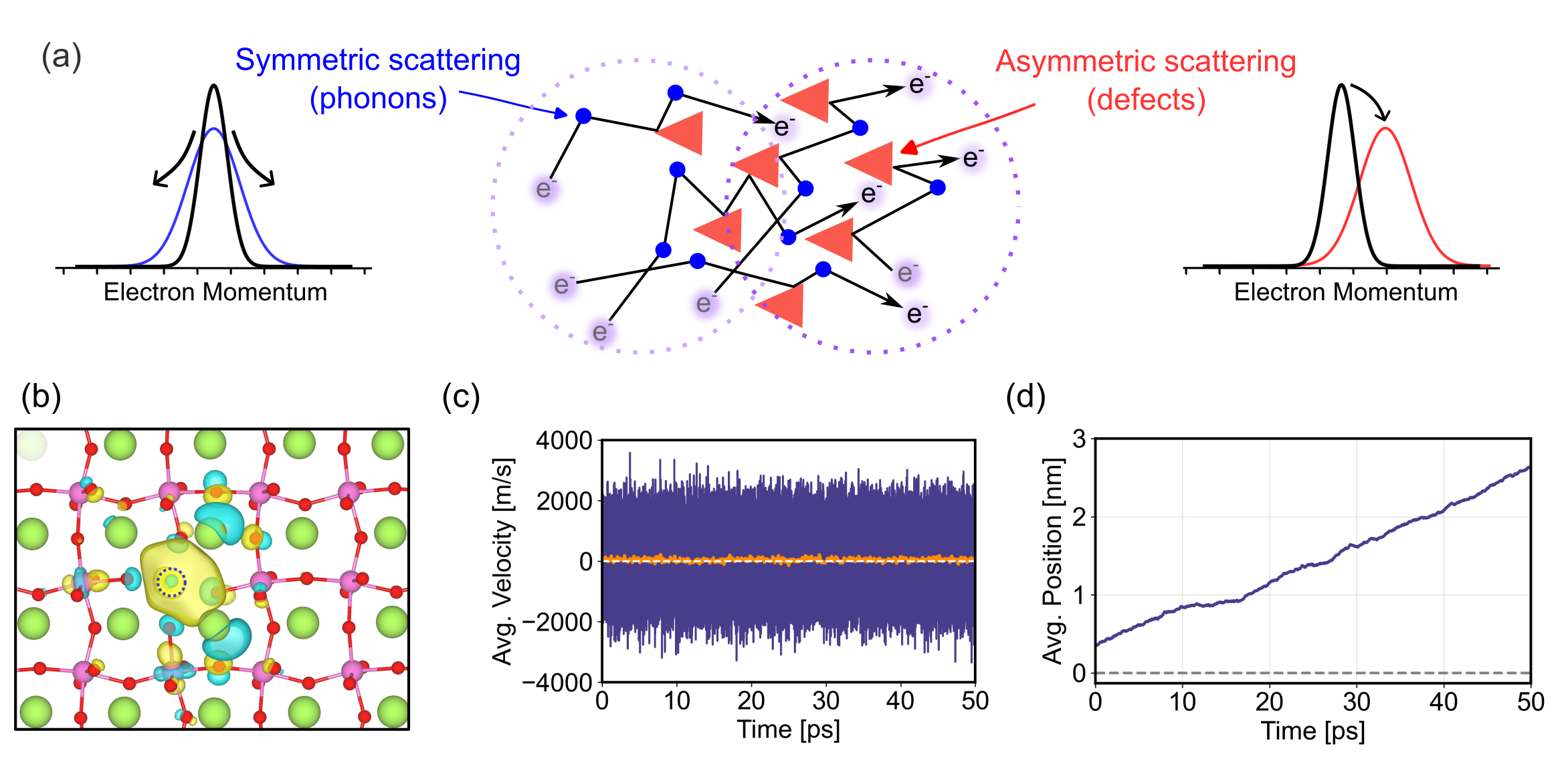}
    \caption{\textbf{Theoretical calculations of asymmetric charge spreading. }
    \textbf{a)}. Schematic illustration of the two main mechanisms of carrier scattering considered in the model. A population of excited carriers with a symmetric velocity distribution (in black) may scatter incoherently from lattice vibrations (blue circles in the middle image), resulting in a symmetric broadening of the distribution (left image). In contrast, scattering from localized defect sites with anisotropic electronic structure (red triangles in the middle image, following the argument in Ref.~\cite{sturman1992}) will result in a non-zero average carrier velocity (right image). The exemplary real-space trajectories in the middle image show how this leads to a net displacement of charges, indicated by the dashed circles before (light purple) and after (dark purple) scattering.  
    \textbf{b)} Wavefunction of an in-gap oxygen vacancy state in BiFeO\textsubscript{3}, with the vacancy location circled, as calculated using DFT (see Methods). The electron density is highly asymmetric and oriented along [111]. 
    \textbf{c)} Average velocity of the carriers. Inclusion of asymmetric scattering leads to a net drift of the charges and a nonzero average velocity $\approx$ 46 m/s (smoothed value in orange). 
    \textbf{d)} Average position of charge carriers as a function of time, showing linear drift calculated using Monte Carlo simulations (see Methods).}
    \label{fig:3}
\end{figure}

We now turn to simulations of the microscopic transport of photoexcited charge carriers. In an asymmetric material such as BiFeO\textsubscript{3}, not only the excitation, but also the microscopic probability of carrier trapping at defects can be asymmetric \cite{belinicher1977photogalvanic}. Defect-related potentials can also cause asymmetric momentum scattering \cite{ruschhaupt_asymmetric_2018}, which can lead to asymmetric carrier responses. In this case, even carriers near the bottom of the conduction band may continue to diffuse asymmetrically. Since this asymmetric momentum scattering can occur as long as photocarriers remain in the conduction band, we hypothesize that this mechanism can explain the long-lived currents observed in our spatiotemporal carrier mapping. Figure \ref{fig:3}a illustrates how asymmetric scattering leads to a persistent net current after multiple scattering events. 
\\

In BiFeO\textsubscript{3}, the predominant defect is the oxygen vacancy, which has an acceptor level 0.6 eV below the band gap \cite{clark2009energy}. To examine the capability of oxygen vacancies to scatter carriers asymmetrically, we conduct first principles calculations (see Methods). We find that the electron density of the sub-conduction band oxygen vacancy state in BiFeO\textsubscript{3} is highly asymmetric, see Fig.\ \ref{fig:3}b. We therefore propose oxygen vacancy trapping/scattering as a likely source of asymmetric charge spreading.   
\\

Carriers also experience incoherent electron-phonon scattering, which tends to symmetrize the momentum distribution, so we must consider whether the momentum asymmetry generated by the defects can cause a prolonged drift in the presence of other scattering processes. We employ a Monte Carlo method (see Methods for details) to simulate the momentum and position of carriers. In the simulations, carrier momenta may be scattered both symmetrically due to electron-phonon interactions, or asymmetrically due to defect scattering, see Fig.~\ref{fig:3}a. 
For the probabilities of symmetric and asymmetric scattering events we use $\tau^{-1}_{\rm{sym}} = \qty{2}{\per\fs}$ and $\tau^{-1}_{\rm{asym}} = \qty{2}{\per\ps}$, respectively. The symmetric scattering time is chosen so that the carrier diffusion calculated from the Monte Carlo matches the observed diffusion rate $\alpha = \qty[per-mode=symbol]{0.06}{\cm\squared\per\s}$. The asymmetric scattering rate is chosen based on the density of oxygen vacancies, as outlined in the Methods.
We calculate the temporal evolution of the average velocity and position, as shown in Figs. \ref{fig:3}c-d. The carriers themselves can have very large velocities in between scattering events, leading to rapid fluctuations even when averaging over the simulated ensemble. However, by averaging also in time, we find that the net carrier velocity is nonzero with a magnitude of $\approx$ \qty{46}{m/s} -- close to the experimentally observed drift velocity. We obtain this value without explicitly applying any constraints on the velocity, either in the initial conditions or during the sampling of scattering processes. These results thus indicate that scattering from defects can indeed play an important role in realizing long-lived asymmetric carrier motion even in the presence of other much more frequent symmetric scattering processes. 
\\

The simulations provide several insights. First, we only obtain a constant net drift velocity when including an asymmetric scattering process, and a larger asymmetric scattering cross-section gives a larger drift velocity. Second, it is necessary to have an initial photogenerated carrier distribution that is out-of-equilibrium, but not necessarily asymmetric. This suggests that the mechanism of defect-mediated asymmetric scattering by itself can explain the occurrence of bulk photovoltaic currents, even without an initial ``kick'' by the ballistic or shift mechanisms. 
\\




\section*{Discussion}


By directly visualizing asymmetric currents in space and time in a non-centrosymmetric material without contacts and interfaces, we established the intrinsic bulk origin of the BPVE. The unexpectedly long persistence of the asymmetric transport ($>$10 ns) -- far beyond the ultrashort timescales associated with conventional shift and ballistic current mechanisms -- revealed a previously unrecognized microscopic mechanism underlying the BPVE. Through Monte Carlo simulations and first-principles calculations, we identified asymmetric momentum scattering by defects as the microscopic origin of this long-lived current. 
We note that the resulting bulk photovoltaic response is of the ballistic type, as it leads to asymmetric carrier populations. Since it is quite distinct from the conventional ballistic current BPVE mechanism in the absence of defects, we refer to it as ``defect-enabled BPVE'' for clarity. We believe that similar mechanisms can play an important role in other materials that display BPVE responses. 
\\

It is quite straightforward to envision how the long-lived asymmetric transport of the ``defect-enabled BPVE'' mechanism can lead to substantial photocurrent generation at the relatively long time and length scales of photovoltaic devices. This is not directly clear for the short-lived asymmetries from shift and ballistic current BPVE mechanisms. Our results also provide a microscopic explanation of the results by Choi \textit{et al.}, who observed unidirectional electronic transport in devices based on BiFeO\textsubscript{3} and suggested that polarization-related asymmetry of impurity potentials could play a role \cite{choi_switchable_2009}. The results are moreover consistent with observations of long-lived photostriction of BiFeO\textsubscript{3} \cite{schick2014localized}, which was tentatively attributed to photocarrier trapping. Finally, we suggest defect engineering as a promising strategy for controlling photovoltaic and photogalvanic responses in polar materials.
\\

\newpage
\section*{Online Methods}

\subsection*{Structural characterization}

We performed Raman scattering experiments using a WITEC R300 model with a 532 nm source at normal incidence to the sample surface (see \ref{fig:ext1}c). An incident power of 1 mW over an estimated spot diameter of 1 $\mu$m was used with a 50$\times$ objective. The integration time was 1 second, and the grating density for the spectrometer was set to 600 per millimeter. We performed the X-ray diffraction study on a Panalytical X'pert Pro diffractometer (Copper K\textsubscript{$\alpha$1}) using a parabolic mirror and a (220)Ge double-pass monochromator on the incident beam side and a PIXcel position-sensitive detector. A $\Phi$ scan was performed for the (110) reflection with a $\chi$ of \ang{45} (see \ref{fig:ext1}d). 

\subsection*{Electrical measurements}

We measured the surface topography and piezoresponse (see \ref{fig:ext1}e) with an MFP-3D Asylum AFM (Asylum Research~-Oxford Instruments). For the in-plane piezoresponse force microscopy measurements, PPP-EFM (Nano sensors; Schaffhausen, Switzerland) tips with a force constant of 0.5 - 9.5 N/m and a resonant frequency of 75 kHz were used. For BPVE measurements, a continuous wave diode laser (Cobolt MLD, 405 nm) was modulated with a square pulse of time period 0.07 s and duty cycle of 50\% from a signal generator. We rotated the polarization of the incident light using a half-wave plate with the crystal placed at normal incidence. A pair of gold electrodes (50 nm) were deposited with a gap of 28 $\mu$m by e-beam evaporation, and the two ends were connected to the input of a lock-in amplifier (MFLI Zurich Instruments) across a high-resistance channel. The sync signal from the signal generator was used as a reference for the lock-in. For measuring the $I-V$ curve, a steady state illumination was provided and at each voltage step and an external bias was applied through a Keithley 2604b sourcemeter. The current was measured in a low-resistance channel of the lock-in after a waiting time of 10 minutes. 
\\

\subsection*{Optical measurements}

\subsubsection*{Absorbance and Transient Photoluminescence}

The absorbance was measured with a Hyperion-2000 system in transmittance mode on a thin flake of the BiFeO\textsubscript{3}  crystal. The transient photoluminescence was measured using a fluorometer from Horiba (Fluorolog 3-11) synchronized with a time-correlated single photon counting unit and a pulsed blue laser for excitation (NanoLED-405L, less than 100 ps pulse width, 1 MHz repetition rate). The detection was done at 430 nm. The instrument response function (IRF) was recorded using a bare glass substrate to scatter the excitation beam.

\subsubsection*{Spatiotemporal microscopy}

\ref{fig:ST} shows a schematic of the spatiotemopral pump-probe setup, where a mode-locked laser ($f_{rep}$ = 76 MHz) generates pulses centered at 1030 nm. Most of the laser output power is used to pump an optical parametric oscillator (OPO) which has a tunable signal output between 1320 and 2000 nm. Using a Lithium Triborate (LBO) crystal, we generate a second harmonic (515 nm) of the fundamental which acts as the pump. A mechanical chopper is used to modulate the pump beam (chopping frequency = 4.37 kHz). The output of the OPO is also frequency-doubled to 770 nm and is used as the probe beam. The pump and probe beams are combined with a dichroic mirror and focused onto the sample plane using a 50$\times$ microscope objective lens (NA: 0.67); both beams have a linear polarization but are cross-polarized with respect to each other. The pump and the probe beam foci were overlapped in the sample plane by iterative (de)collimation and beam profiling through knife edge scans, resulting in a pump spot $1/e$ value of 0.34 $\upmu$m and probe spot of 0.44 $\upmu$m at the same $z$ position.  The fluence is calculated as $F = P/ (f_{rep} \cdot \pi r^{2}_{1/e})$.  For spatial scans, the probe beams were steered through a scanning mirror. A variable delay line is employed to acquire pump-probe time delay dependent transient reflection measurements. The transient signal from the sample is detected by a Si photodiode and the signal is demodulated using a lock-in amplifier (Zurich Instruments MFLI) corresponding to the chopping frequency. The pump and probe fluences were fixed at 8.7 J/m\textsuperscript{2} and 7.2 J/m\textsuperscript{2} respectively for all reflectivity measurements, unless specified. 
\\

\subsection*{Phenomenological population dynamics model}
To model the spatiotemporal evolution of the photoexcited population, we describe the dynamics using drift-diffusion–decay equations for two populations of photoexcited carriers. The population $n_1$ undergoes diffusion and decay, see Eq.~\ref{eq:N1}, while the population $n_2$ undergoes drift in addition to diffusion and decay, see Eq.~\ref{eq:N2}). The total population dynamics are obtained from the sum $n_1 + n_2$. These populations evolve according to the following equations, where $\alpha$ is the diffusion coefficient, $\tau$ is the effective photoexcited carrier lifetime, and $v_d$ is the drift velocity, where a negative $v_d$ corresponds to motion along the $-x$ direction:

\begin{equation}
\frac{\partial n_1}{\partial t}
= \alpha \nabla^2 n_1 - \frac{1}{\tau} n_1 ,
\label{eq:N1}
\end{equation}

\begin{equation}
\frac{\partial n_2}{\partial t}
= \alpha \nabla^2 n_2
- v_d \frac{\partial n_2}{\partial x}
- \frac{1}{\tau} n_2 .
\label{eq:N2}
\end{equation}

In our simulation, we mimic the excitation by a pump pulse by initializing both populations with a Gaussian spatial profile with a width of $\sigma_{\mathrm{pu}} = 0.24\,\upmu$m. Because the time interval between consecutive laser pulses ($13\,\text{ns}$, corresponding to a repetition rate $f_{\mathrm{rep}} = 76\,\text{MHz}$) is shorter than the characteristic decay time of the population, the photoexcited carriers do not fully relax back to valence band before the arrival of the subsequent pulse. The population dynamics are therefore simulated by sequentially injecting Gaussian source terms at time intervals of $1/f_{\mathrm{rep}}$ into the evolving population distribution. Each excitation pulse adds a Gaussian population profile to the existing partially decayed distribution. This procedure is repeated for a total number of pulses determined by the laser repetition rate and the pump modulation frequency $f_{\mathrm{mod}}$, given by $N = \frac{f_{\mathrm{rep}}}{2 f_{\mathrm{mod}}}$ for a $50\%$ duty cycle. For the experimental parameters used here ($f_{\mathrm{rep}} = 76\,\text{MHz}$ and $f_{\mathrm{mod}} = 4.37\,\text{kHz}$), this corresponds to approximately $8700$ excitation pulses. The final simulated population distribution is convolved with the probe beam profile, which has a width of $\sigma_{\mathrm{pr}} = 0.31\,\upmu$m. The resulting profile is shown in Fig.~\ref{fig:Data_Sim}c. 
\\

\subsection*{First principles calculation}

Plane-wave pseudopotential density functional theory calculations were performed using the \textsc{Quantum Espresso} code \cite{giannozzi2009quantum,giannozzi2017advanced}. Designed nonlocal norm-conserving pseudopotentials were generated using the OPIUM package \cite{rappe1990optimized, Ramer99p12471}.
The generalized gradient exchange-correlation functional of Perdew-Burke-Ernzerhof was used with Hubbard $U$ ($U = \qty{5.0}{\eV}$) correction for the Fe 3$d$ orbitals \cite{perdew1996generalized}.
A Monkhorst-Pack reciprocal space sampling was adopted, with a spacing of at most 0.22~\AA$^{-1}$\ between the neighboring k-points. The plane-wave cutoff was set to 50 Ry and was converged with self-consistency convergence criteria of $10^{-9}$ eV/unit cell.
A pristine 270-atom supercell was constructed from a fully relaxed rhombohedral BiFeO\textsubscript{3} (spacegroup \#161 R3c) bulk crystal structure. An oxygen vacancy was modeled by removing a single oxygen atom from this supercell. The lattice parameters of the supercell were fixed while the atomic positions were allowed to relax. The defect cell had a single in-gap defect state, and the wave function of this state is presented in Fig.\ref{fig:3}b.

\subsection*{Kinetic Monte-Carlo Simulation}

A Kinetic Monte Carlo algorithm was used to model carrier interactions with defects and phonons. We followed the algorithm described in Ref.~\cite{jacoboni_monte_1983} to model electron-phonon scattering of velocities and used classical molecular dynamics to propagate the trajectories of carriers in real space.
Two different scattering mechanisms were modeled: 1) symmetric electron-acoustic phonon scattering in the lattice system of BiFeO\textsubscript{3}, and 2) asymmetric electron-acoustic phonon scattering at defect sites modulated by the asymmetric defect dipole.

In the simulation, a carrier with momentum $\bm{k}$ is scattered symmetrically by acoustic phonons to momentum $\bm{k} + \bm{q}$ with a probability per unit time of $P(k, q)dq =\frac{m^* \mathcal{E}_{1}^{2}}{4 \pi \rho v_{s} \hbar^{2} k}\binom{N_q}{N_{q} + 1} q^{2} dq$, where the upper and lower terms correspond to absorption and emission of a phonon. Here, $k$ and $q$ are the magnitudes of $\bm{k}$ and $\bm{q}$, and the simulation parameters chosen are the carrier effective mass $m^* = \num{0.393} m_e$ (where $m_e$ is the electron rest mass), the electron-phonon deformation potential $\mathcal{E}_{1} = \qty{15}{\eV}$, the density of BiFeO\textsubscript{3} $\rho = \qty{8.1e3}{\kg\per\m\cubed}$, and the sound velocity $v_s = \qty{4.32e3}{\m\per\s}$. $N_{q}$ is the Bose-Einstein equilibrium occupation number of phonons with wavevector $\bm{q}$.
In a symmetric scattering event, the magnitude $q$ is first sampled based on the equation above, and the direction is chosen uniformly while constrained by energy and momentum conservation. These symmetric scattering events occur randomly with a rate of $\tau_{\rm{sym}}^{-1} = \qty{2}{\per\fs}$, and the simulation timestep is \qty{0.01}{\fs}.

The asymmetric scattering follows a similar form, but in this case is modulated by a directional component. This makes the electronic subsystem Hamiltonian non-Hermitian. Here, the probability for scattering from momentum $\bm{k}$ to $\bm{k'}$ is $P(\bm{k}, \bm{k'}) = \frac{\pi q \mathcal{E}_{1}^{2}}{\rho v_{s}}\binom{N_{q}}{N_{q}+1} (1 + \eta \bm{\hat{d}} \cdot \bm{\hat{q}}) \delta \left[ \epsilon(\bm{k'}) - \epsilon(\bm{k}) \mp \hbar q v_{s} \right]$.
The additional simulation parameters are the orientation of the asymmetry $\bm{\hat{d}} = \bm{\hat{x}}$, the degree of symmetry breaking $\eta = 1$ (the asymmetric scattering would reduce back to symmetric case if $\eta = 0$).
We assume that these asymmetric scattering events occur from encounters with defects, and so we approximate the rate of asymmetric scattering by $\tau_{\rm{asym}}^{-1} \approx n_d \sigma_d v_F$, for defect density $n_d$, defect cross section $\sigma_d$ and Fermi velocity $v_F$. Oxygen vacancies at \qty{1}{\percent} give a defect density of \qty{4.7}{\per\nm\cubed} and the Fermi velocity is $v_F = \qty{1.5e6}{\m\per\s}$.
The scattering cross section is difficult to find in the literature, but we can take an approximate magnitude from electron-phonon scattering. Several sources give e-ph cross sections in the range of \qtyrange{0.01}{0.1}{\angstrom\squared}. If we take a value at the lower end of this range, we find the corresponding scattering lifetime to be on the order of $\tau_{\rm{asym}}^{-1} = \qty{2}{\per\ps}$.
A total of \num{30720} carriers are simulated, and the trajectory for each carrier is simulated for \qty{50}{\ps}.

\newpage
\printbibliography

@book{sturman2021photovoltaic,
  title={Photovoltaic and photo-refractive effects in noncentrosymmetric materials},
  author={Sturman, Boris and Fridkin, Vladimir},
  year={2021},
  publisher={Routledge}
}

@article{spanier2016power,
	title = {Power conversion efficiency exceeding the {Shockley}–{Queisser} limit in a ferroelectric insulator},
	volume = {10},
	number = {9},
	journal = {Nature Photonics},
	author = {Spanier, Jonathan E. and Fridkin, Vladimir M. and Rappe, Andrew M. and Akbashev, Andrew R. and Polemi, Alessia and Qi, Yubo and Gu, Zongquan and Young, Steve M. and Hawley, Christopher J. and Imbrenda, Dominic and Xiao, Geoffrey and Bennett-Jackson, Andrew L. and Johnson, Craig L.},
	year = {2016},
	pages = {611--616}
}

@article{pusch2023energy,
  title={Energy conversion efficiency of the bulk photovoltaic effect},
  author={Pusch, Andreas and R{\"o}mer, Udo and Culcer, Dimitrie and Ekins-Daukes, Nicholas J},
  journal={PRX Energy},
  volume={2},
  number={1},
  pages={013006},
  year={2023},
  publisher={APS}
}

@article{belinicher1980photogalvanic,
  title = {The photogalvanic effect in media lacking a center of symmetry},
  volume = {23},
  ISSN = {0038-5670},
  number = {3},
  journal = {Soviet Physics Uspekhi},
  publisher = {IOP Publishing},
  author = {Belinicher,  V I and Sturman,  B I},
  year = {1980},
  month = mar,
  pages = {199–223}
}

@article{belinicher1988relation,
	title = {The relation between shift and ballistic currents in the theory of photogalvanic effect},
	volume = {83},
	number = {1},
	journal = {Ferroelectrics},
	author = {Belinicher, V. I. and Sturman, B. I.},
	year = {1988},
	pages = {29--34}
}

@book{sturman1992,
  address={Philadelphia},
  edition={1},
  series={Ferroelectricity and Related Phenomena},
  title={The Photovoltaic and Photorefractive Effects in Noncentrosymmetric Materials},
  ISBN={2-88124-498-X},
  note={Translated from the Russian by J.E.S. Bradley},
  publisher={Gordon and Breach},
  author={Sturman, Boris I. and Fridkin, Vladimir M.},
  year={1992},
  collection={Ferroelectricity and Related Phenomena},
  language={en}
}

@article{sipe2000second,
  title={Second-order optical response in semiconductors},
  author={Sipe, JE and Shkrebtii, AI},
  journal={Physical Review B},
  volume={61},
  number={8},
  pages={5337},
  year={2000},
  publisher={APS}
}

@article{dai2023recent,
  title={Recent progress in the theory of bulk photovoltaic effect},
  author={Dai, Zhenbang and Rappe, Andrew M},
  journal={Chemical Physics Reviews},
  volume={4},
  number={1},
  year={2023},
  publisher={AIP Publishing}
}

@article{vonBaltz1981theory,
	title = {Theory of the bulk photovoltaic effect in pure crystals},
	volume = {23},
	issn = {0163-1829},
	number = {10},
	urldate = {2022-04-27},
	journal = {Physical Review B},
	author = {von Baltz, Ralph and Kraut, Wolfgang},
	month = may,
	year = {1981},
	pages = {5590--5596}
}

@article{Belinicher82p649,
  title={Kinetic theory of the displacement photovoltaic effect in piezoelectrics},
  volume={83},
  journal={Zh. Eksp. Teor. Fiz.},
  author={Belinicher, V I and Ivchenko, E L and Sturman, B I},
  year={1982},
  pages={649–661}
}

@article{uzan2024observation,
  title={Observation of interband Berry phase in laser-driven crystals},
  author={Uzan-Narovlansky, Ayelet J and Faeyrman, Lior and Brown, Graham G and Shames, Sergei and Narovlansky, Vladimir and Xiao, Jiewen and Arusi-Parpar, Talya and Kneller, Omer and Bruner, Barry D and Smirnova, Olga and others},
  journal={Nature},
  volume={626},
  number={7997},
  pages={66--71},
  year={2024},
  publisher={Nature Publishing Group UK London}
}

@article{Nagaosa17p1603345,
  title = {Concept of Quantum Geometry in Optoelectronic Processes in Solids: Application to Solar Cells},
  volume = {29},
  ISSN = {1521-4095},
  number = {25},
  pages = {1603345},
  journal = {Advanced Materials},
  publisher = {Wiley},
  author = {Nagaosa,  Naoto and Morimoto,  Takahiro},
  year = {2017}
}

@article{akamatsu2021van,
  title={A van der Waals interface that creates in-plane polarization and a spontaneous photovoltaic effect},
  author={Akamatsu, Takatoshi and Ideue, Toshiya and Zhou, Ling and Dong, Yu and Kitamura, Sota and Yoshii, Mao and Yang, Dongyang and Onga, Masaru and Nakagawa, Yuji and Watanabe, Kenji and Taniguchi, Takashi and Laurienzo, Joseph and Huang, Junwei and Ye, Ziliang and Morimoto, Takahiro and Yuan, Hongtao and Iwasa, Yoshihiro},
  journal={Science},
  volume={372},
  number={6537},
  pages={68--72},
  year={2021},
  publisher={American Association for the Advancement of Science}
}

@article{krishna2025terahertz,
  title={Terahertz photocurrent probe of quantum geometry and interactions in magic-angle twisted bilayer graphene},
  author = {Krishna Kumar, Roshan and Li, Geng and Bertini, Riccardo and Chaudhary, Swati and Nowakowski, Krystian and Park, Jeong Min and Castilla, Sebastian and Zhan, Zhen and Pantale{\'o}n, Pierre A. and Agarwal, Hitesh and Batlle-Porro, Sergi and Icking, Eike and Ceccanti, Matteo and Reserbat-Plantey, Antoine and Piccinini, Giulia and Barrier, Julien and Khestanova, Ekaterina and Taniguchi, Takashi and Watanabe, Kenji and Stampfer, Christoph and Refael, Gil and Guinea, Francisco and Jarillo-Herrero, Pablo and Song, Justin C. W. and Stepanov, Petr and Lewandowski, Cyprian and Koppens, Frank H. L.},
  journal={Nature Materials},
  pages={1--8},
  year={2025},
  publisher={Nature Publishing Group UK London}
}

@article{yang2024driving,
  title={Driving and detecting topological phase transition in noncentrosymmetric systems via an all-optical approach},
  author={Yang, Mengtong and Miao, Xiaoyan and Li, Si and Zhou, Jian and Zhang, Chunmei},
  journal={Physical Review B},
  volume={109},
  number={12},
  pages={125101},
  year={2024},
  publisher={APS}
}

@article{ahn2020low,
  title={Low-frequency divergence and quantum geometry of the bulk photovoltaic effect in topological semimetals},
  author={Ahn, Junyeong and Guo, Guang-Yu and Nagaosa, Naoto},
  journal={Physical Review X},
  volume={10},
  number={4},
  pages={041041},
  year={2020},
  publisher={APS}
}

@article{liang2023strong,
  title={Strong bulk photovoltaic effect in engineered edge-embedded van der Waals structures},
  author = {Liang, Zihan and Zhou, Xin and Zhang, Le and Yu, Xiang-Long and Lv, Yan and Song, Xuefen and Zhou, Yongheng and Wang, Han and Wang, Shuo and Wang, Taihong and Shum, Perry Ping and He, Qian and Liu, Yanjun and Zhu, Chao and Wang, Lin and Chen, Xiaolong},
  journal={Nature Communications},
  volume={14},
  number={1},
  pages={4230},
  year={2023},
  publisher={Nature Publishing Group UK London}
}

@article{zeng2024dual,
  title={Dual polarization-enabled ultrafast bulk photovoltaic response in van der Waals heterostructures},
  author = {Zeng, Zhouxiaosong and Tian, Zhiqiang and Wang, Yufan and Ge, Cuihuan and Strau{\ss}, Fabian and Braun, Kai and Michel, Patrick and Huang, Lanyu and Liu, Guixian and Li, Dong and Scheele, Marcus and Chen, Mingxing and Pan, Anlian and Wang, Xiao},
  journal={Nature Communications},
  volume={15},
  number={1},
  pages={5355},
  year={2024},
  publisher={Nature Publishing Group UK London}
}

@article{tan2016enhancement,
  title={Enhancement of the bulk photovoltaic effect in topological insulators},
  author={Tan, Liang Z and Rappe, Andrew M},
  journal={Physical review letters},
  volume={116},
  number={23},
  pages={237402},
  year={2016},
  publisher={APS}
}

@article{osterhoudt2019colossal,
  title={Colossal mid-infrared bulk photovoltaic effect in a type-I Weyl semimetal},
  author = {Osterhoudt, Gavin B. and Diebel, Laura K. and Gray, Mason J. and Yang, Xu and Stanco, John and Huang, Xiangwei and Shen, Bing and Ni, Ni and Moll, Philip J. W. and Ran, Ying and Burch, Kenneth S.},
  journal={Nature Materials},
  volume={18},
  number={5},
  pages={471--475},
  year={2019},
  publisher={Nature Publishing Group UK London}
}

@article{wang2019robust,
  title={Robust edge photocurrent response on layered type II Weyl semimetal WTe2},
  author = {Wang, Qinsheng and Zheng, Jingchuan and He, Yuan and Cao, Jin and Liu, Xin and Wang, Maoyuan and Ma, Junchao and Lai, Jiawei and Lu, Hong and Jia, Shuang and Yan, Dayu and Shi, Youguo and Duan, Junxi and Han, Junfeng and Xiao, Wende and Chen, Jian-Hao and Sun, Kai and Yao, Yugui and Sun, Dong},
  journal={Nature Communications},
  volume={10},
  number={1},
  pages={5736},
  year={2019},
  publisher={Nature Publishing Group UK London}
}

@article{de2017quantized,
  title={Quantized circular photogalvanic effect in Weyl semimetals},
  author={De Juan, Fernando and Grushin, Adolfo G and Morimoto, Takahiro and Moore, Joel E},
  journal={Nature Communications},
  volume={8},
  number={1},
  pages={15995},
  year={2017},
  publisher={Nature Publishing Group UK London}
}

@article{ji2019spatially,
  title={Spatially dispersive circular photogalvanic effect in a Weyl semimetal},
  author = {Ji, Zhurun and Liu, Gerui and Addison, Zachariah and Liu, Wenjing and Yu, Peng and Gao, Heng and Liu, Zheng and Rappe, Andrew M. and Kane, Charles L. and Mele, Eugene J. and Agarwal, Ritesh},
  journal={Nature Materials},
  volume={18},
  number={9},
  pages={955--962},
  year={2019},
  publisher={Nature Publishing Group UK London}
}

@article{xie2024surface,
  title={Surface photogalvanic effect in Ag2Te},
  author = {Xie, Xiaoyi and Leng, Pengliang and Ding, Zhenyu and Yang, Jinshan and Yan, Jingyi and Zhou, Junchen and Li, Zihan and Ai, Linfeng and Cao, Xiangyu and Jia, Zehao and Zhang, Yuda and Zhao, Minhao and Zhu, Wenguang and Gao, Yang and Dong, Shaoming and Xiu, Faxian},
  journal={Nature Communications},
  volume={15},
  number={1},
  pages={5651},
  year={2024},
  publisher={Nature Publishing Group UK London}
}

@article{sturman2020ballistic,
  title={Ballistic and shift currents in the bulk photovoltaic effect theory},
  author={Sturman, Boris Itskhakovich},
  journal={Physics-Uspekhi},
  volume={63},
  number={4},
  pages={407},
  year={2020},
  publisher={IOP Publishing}
}

@article{dai2021phonon,
  title={Phonon-assisted ballistic current from first-principles calculations},
  author={Dai, Zhenbang and Schankler, Aaron M and Gao, Lingyuan and Tan, Liang Z and Rappe, Andrew M},
  journal={Physical review letters},
  volume={126},
  number={17},
  pages={177403},
  year={2021},
  publisher={APS}
}

@article{gu2017mesoscopic,
  title={Mesoscopic free path of nonthermalized photogenerated carriers in a ferroelectric insulator},
  author={Gu, Zongquan and Imbrenda, Dominic and Bennett-Jackson, Andrew L and Falmbigl, Matthias and Podpirka, Adrian and Parker, Thomas C and Shreiber, Daniel and Ivill, Mathew P and Fridkin, Vladimir M and Spanier, Jonathan E},
  journal={Physical Review Letters},
  volume={118},
  number={9},
  pages={096601},
  year={2017},
  publisher={APS}
}

@article{burger2019direct,
  title={Direct observation of shift and ballistic photovoltaic currents},
  author={Burger, Aaron M and Agarwal, Radhe and Aprelev, Alexey and Schruba, Edward and Gutierrez-Perez, Alejandro and Fridkin, Vladimir M and Spanier, Jonathan E},
  journal={Science Advances},
  volume={5},
  number={1},
  pages={eaau5588},
  year={2019},
  publisher={American Association for the Advancement of Science}
}

@article{laman2005ultrafast,
  title={Ultrafast shift and injection currents observed in wurtzite semiconductors via emitted terahertz radiation},
  author={Laman, N and Bieler, M and Van Driel, HM},
  journal={Journal of Applied Physics},
  volume={98},
  number={10},
  year={2005},
  publisher={AIP Publishing}
}

@article{schick2014localized,
  title={Localized excited charge carriers generate ultrafast inhomogeneous strain in the multiferroic {BiFeO$_3$}},
  author={Schick, Daniel and Herzog, Marc and Wen, Haidan and Chen, Pice and Adamo, Carolina and Gaal, Peter and Schlom, Darrell G and Evans, Paul G and Li, Yuelin and Bargheer, Matias},
  journal={Physical Review Letters},
  volume={112},
  number={9},
  pages={097602},
  year={2014},
  publisher={APS}
}

@article{Vazquez2024,
author = {Vazquez, Guillermo D.Brinatti and Morganti, Giulia Lo Gerfo and Block, Alexander and van Hulst, Niek F. and Liebel, Matz and Tielrooij, Klaas Jan},
doi = {10.1002/aelm.202300584},
issn = {2199160X},
journal = {Advanced Electronic Materials},
number = {2},
pages = {1--16},
title = {{Spatiotemporal Microscopy: Shining Light on Transport Phenomena}},
volume = {10},
year = {2024}
}

@article{lebeugle2007room,
  title={Room-temperature coexistence of large electric polarization and magnetic order in {BiFeO$_3$} single crystals},
  author={Lebeugle, Delphine and Colson, Doroth{\'e}e and Forget, Anne and Viret, Michel and Bonville, Pierre and Marucco, Jean-Francis and Fusil, Stephane},
  journal={Physical Review B},
  volume={76},
  number={2},
  pages={024116},
  year={2007},
  publisher={APS}
}

@article{belinicher1977photogalvanic,
  title={Photogalvanic effect in a crystal with polar axis},
  author={Belinicher, VI and Malinovskiǐ, VK and Sturman, BI},
  journal={Soviet Journal of Experimental and Theoretical Physics},
  volume={46},
  pages={362},
  year={1977}
}

@article{ruschhaupt_asymmetric_2018,
	title = {Asymmetric scattering by non-{Hermitian} potentials},
	volume = {120},
	issn = {0295-5075},
	number = {2},
	journal = {Europhysics Letters},
	author = {Ruschhaupt, A. and Dowdall, T. and Simón, M. A. and Muga, J. G.},
	year = {2018},
	pages = {20001}
}

@article{clark2009energy,
  title={Energy levels of oxygen vacancies in BiFeO3 by screened exchange},
  author={Clark, SJ and Robertson, J},
  journal={Applied Physics Letters},
  volume={94},
  number={2},
  year={2009},
  publisher={AIP Publishing}
}

@article{choi_switchable_2009,
	title = {Switchable {Ferroelectric} {Diode} and {Photovoltaic} {Effect} in {BiFeO$_3$}},
	volume = {324},
	number = {5923},
	journal = {Science},
	author = {Choi, T. and Lee, S. and Choi, Y. J. and Kiryukhin, V. and Cheong, S.-W.},
	year = {2009},
	publisher = {American Association for the Advancement of Science},
	pages = {63--66}
}

@article{giannozzi2009quantum,
  title={QUANTUM ESPRESSO: a modular and open-source software project for quantumsimulations of materials},
  author={Giannozzi, Paolo and Baroni, Stefano and Bonini, Nicola and Calandra, Matteo and Car, Roberto and Cavazzoni, Carlo and Ceresoli, Davide and Chiarotti, Guido L and Cococcioni, Matteo and Dabo, Ismaila and others},
  journal={Journal of physics: Condensed matter},
  volume={21},
  number={39},
  pages={395502},
  year={2009},
  publisher={IOP Publishing}
}

@article{giannozzi2017advanced,
  title={Advanced capabilities for materials modelling with Quantum ESPRESSO},
  author={Giannozzi, Paolo and Andreussi, Oliviero and Brumme, Thomas and Bunau, Oana and Nardelli, M Buongiorno and Calandra, Matteo and Car, Roberto and Cavazzoni, Carlo and Ceresoli, Davide and Cococcioni, Matteo and others},
  journal={Journal of physics: Condensed matter},
  volume={29},
  number={46},
  pages={465901},
  year={2017},
  publisher={IOP Publishing}
}

@article{rappe1990optimized,
  title={Optimized pseudopotentials},
  author={Rappe, Andrew M and Rabe, Karin M and Kaxiras, Efthimios and Joannopoulos, JD},
  journal={Physical Review B},
  volume={41},
  number={2},
  pages={1227},
  year={1990},
  publisher={APS}
}

@article{Ramer99p12471,
  title = {Designed nonlocal pseudopotentials for enhanced transferability},
  volume = {59},
  ISSN = {1095-3795},
  number = {19},
  journal = {Physical Review B},
  publisher = {American Physical Society (APS)},
  author = {Ramer,  Nicholas J. and Rappe,  Andrew M.},
  year = {1999},
  pages = {12471–12478}
}

@article{perdew1996generalized,
  title={Generalized gradient approximation for the exchange-correlation hole of a many-electron system},
  author={Perdew, John P and Burke, Kieron and Wang, Yue},
  journal={Physical review B},
  volume={54},
  number={23},
  pages={16533},
  year={1996},
  publisher={APS}
}

@article{varghese2023,
    author = {Varghese, Sebin and Mehew, Jake Dudley and Block, Alexander and Reig, David Saleta and Woźniak, Paweł and Farris, Roberta and Zanolli, Zeila and Ordejón, Pablo and Verstraete, Matthieu J. and van Hulst, Niek F. and Tielrooij, Klaas-Jan},
    title = {A pre-time-zero spatiotemporal microscopy technique for the ultrasensitive determination of the thermal diffusivity of thin films},
    journal = {Review of Scientific Instruments},
    volume = {94},
    number = {3},
    pages = {034903},
    year = {2023}
}

@article{moubah2012photoluminescence,
  title={Photoluminescence investigation of defects and optical band gap in multiferroic BiFeO3 single crystals},
  author={Moubah, Reda and Schmerber, Guy and Rousseau, Olivier and Colson, Doroth{\'e}e and Viret, Michel},
  journal={Applied Physics Express},
  volume={5},
  number={3},
  pages={035802},
  year={2012},
  publisher={IOP Publishing}
}

@article{jacoboni_monte_1983,
	title = {The {Monte} {Carlo} method for the solution of charge transport in semiconductors with applications to covalent materials},
	volume = {55},
	number = {3},
	urldate = {2024-02-20},
	journal = {Rev. Mod. Phys.},
	author = {Jacoboni, Carlo and Reggiani, Lino},
	year = {1983},
	publisher = {American Physical Society},
	pages = {645--705}
}

@article{zheng2026photothermal,
  title={Photothermal effects control ultrafast charge transport in titanium carbide MXenes},
  author={Zheng, Wenhao and Ramsden, Hugh and Ippolito, Stefano and van Hemert, Max and Zhang, Danzhen and Zhang, Teng and Li, Dongqi and Wen, Guanzhao and Geuchies, Jaco J and Yu, Minghao and others},
  journal={Nature communications},
  year={2026},
  publisher={Nature Publishing Group UK London}
}

\newpage 
\subsection*{Acknowledgments}
The authors acknowledge Dr. Rafael Sanchez (INAM, UJI), Dr. Kumara Cordero (ICN2), Dr. Jessica Padilla (ICN2)  for assisting with transient photoluminescence, piezoresponse force microscopy and X-ray diffraction measurements, respectively. S.G. acknowledges the PREBIST Cofund grant. This project has received funding from the European Union’s Horizon 2020 research and innovation programme under the Marie Skłodowska-Curie grant agreement No. 754558. S.V. acknowledges the support of the Spanish Ministry of Economy through FPI-SO2018. G.C. acknowledges financial support from the Catalan government (grant number 2021 SGR 0129), and from the Spanish Research Agency (Agencia Estatal de Investigación), project number PID2023-148673NB-I00. K.J.T. acknowledges funding from the European Union’s Horizon 2020 research and innovation program under Grant Agreement No. 101125457 (ERC CoG ``EQUATE'') and Spanish MCIN/AEI project PID2022-142730NB-I00 ``HYDROPTO'').
All research at ICN2 is supported by a Severo Ochoa Grant CEX2021-001214-S. 
A.M.S., K.M., X.X., and A.M.R. acknowledge support for theoretical modeling of quantum scattering processes from the U.S. Department of Energy, Office of Science, Basic Energy Sciences, under Award No. DE-SC0024942. Computational support was provided by the National Energy Research Scientific Computing Center (NERSC), a U.S. Department of Energy, Office of Science User Facility located at Lawrence Berkeley National Laboratory, operated under Contract No. DE-AC02-05CH11231. K.M. acknowledges the JSPS Overseas Research Fellowship.

\subsection*{Competing interests}

The authors declare no competing interests. 

\subsection*{Data availability}

The data generated and analysed supporting the findings of this work are available from the corresponding author upon reasonable request.

\subsection*{Author contributions}

G.C. and K.J.T. initiated and coordinated the work; S.G performed the optical, structural and electrical measurements on the single crystal provided by M.V.; S.G. and S.V. performed the spatiotemporal measurements on the single crystal and analyzed the data under the supervision of G.C. and K.J.T; A.M.S, X.X, and K.M. developed and implemented the theoretical models under the supervision of A.M.R; K.J.T, A.M.S, K.M, G.C, and S.G. co-wrote the manuscript, with input from all authors. All authors participated in the discussions of this work. 

\newpage
\section*{Extended Data Figures}

\captionsetup[figure]{format=plain, indention=0pt}
\renewcommand{\figurename}{}
\renewcommand{\thefigure}{Extended Data Figure \arabic{figure}}
\setcounter{figure}{0}

\begin{figure}[H]
    \centering
    \includegraphics[width=\textwidth]{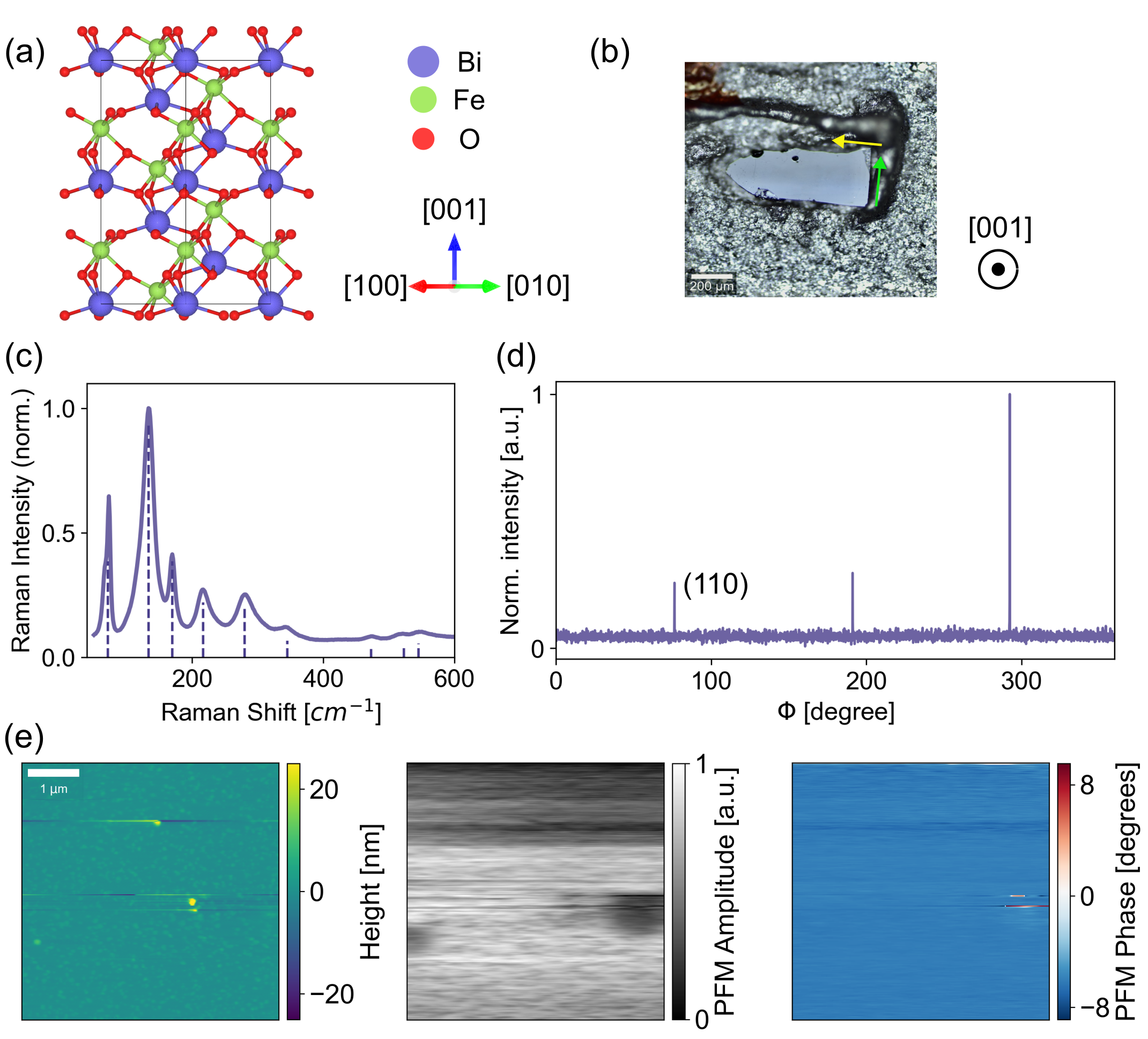}
    \caption{\textbf{Structural and electrical characterization of the BiFeO\textsubscript{3} single crystal} 
    \textbf{a)} The crystal structure of ferroelectric BiFeO\textsubscript{3} with a projection vector of $\langle 110 \rangle$ and upward vector $\langle 001 \rangle$.
    \textbf{b)} Optical image of the single crystal. The green arrow is parallel to [1-10], while the yellow arrow is parallel to [110], the in-plane component of the [111] oriented ferroelectric polarization. The surface normal is parallel to [001].
    \textbf{c)} Raman spectrum of the single crystal, showing all the 9 possible transverse optical phonon modes allowed for R3c symmetry as expected for a monodomain crystal in $Z(XY)\overline{Z}$ configuration (Porto's notation).  
    \textbf{d)} $\Phi$ scan in an X-ray diffraction experiment, using the (110) reflection with a $\chi$ of 45 degrees. Both Raman and X-ray diffraction confirm the single crystal state of the sample \textbf{e)} Topography of the sample (left panel), Amplitude (middle panel) and phase (right panel) in in-plane piezoresponse force microscopy. The uniform phase image confirms a single orientation of the ferroelectric polarization in the single crystal.}
    \label{fig:ext1}
\end{figure}

\begin{figure}[H]
    \centering
    \includegraphics[width=0.9\textwidth]{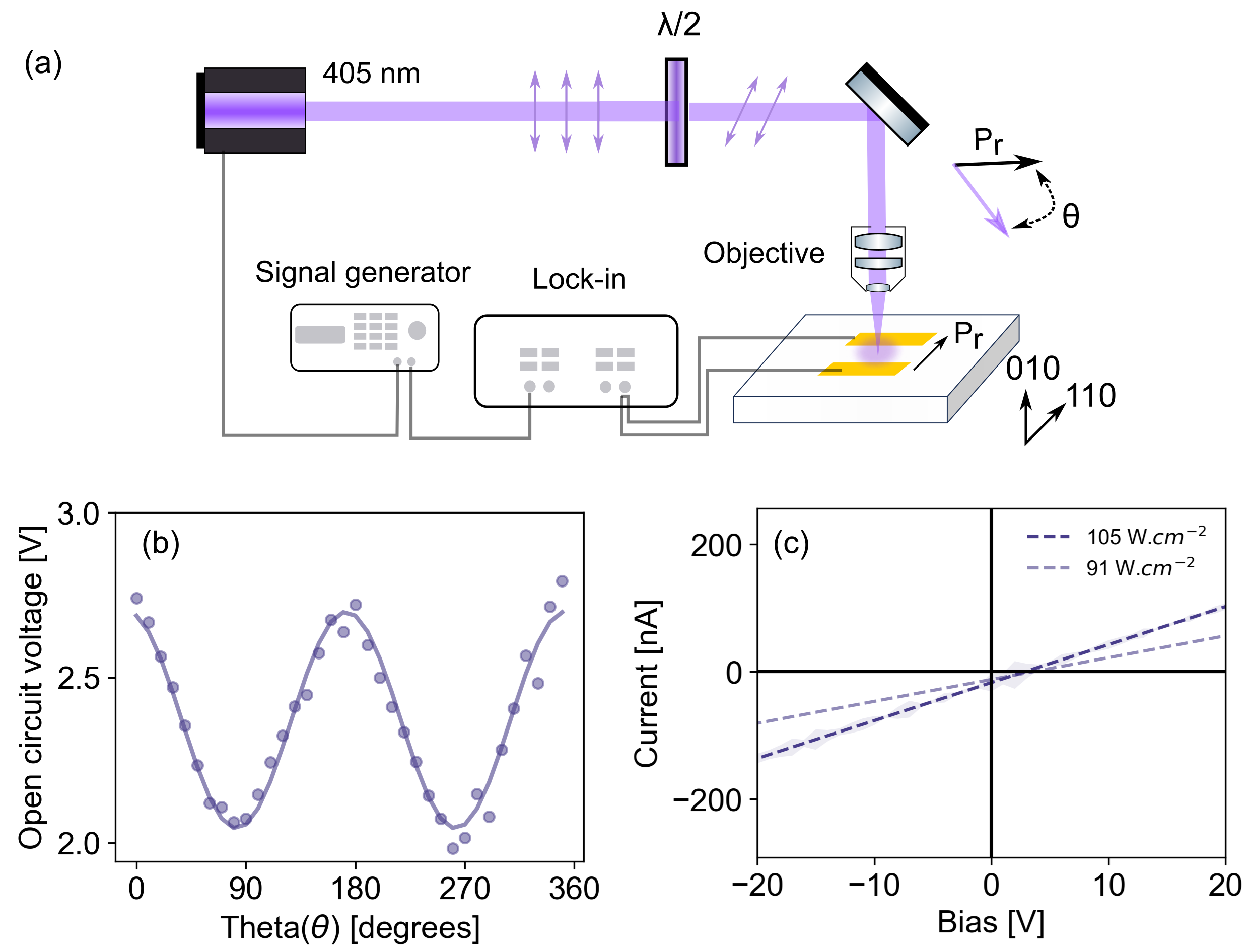}
    \caption{\textbf{Optoelectronic measurement of the bulk photovoltaic effect} 
    \textbf{a)} Schematic of the optoelectronic setup used for measuring the photocurrent due to the BPVE. 
    \textbf{b)} The variation in photovoltage across the crystal when incident light polarization is rotated with respect to the crystal polarization, a hallmark of BPVE. The photovoltage is maximum when light polarization is parallel or antiparallel to the crystal polarization and minimum when perpendicular. \textbf{b}. I-V curve of the single crystal under illumination with a short-circuit current of $\sim$17.7 nA and an open-circuit voltage of 3.1 V, which is slightly larger than the bandgap and therefore indicative of the bulk origin of the photovoltage.}
    \label{fig:ext2}
\end{figure}

\begin{figure}[h]
    \centering
    \includegraphics[width=0.7\linewidth]{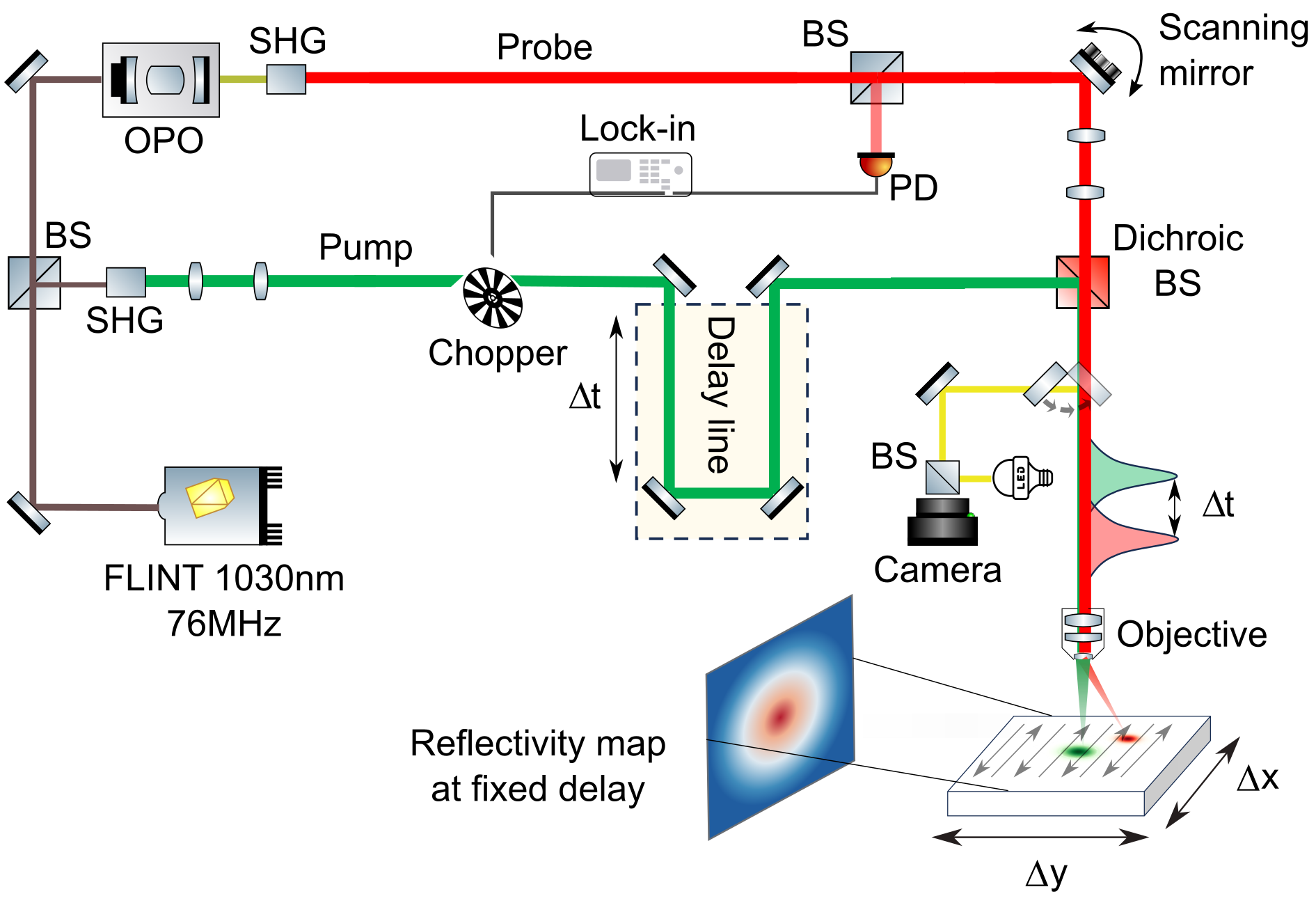}
    \caption{\textbf{Spatiotemporal pump-probe microscopy setup}. A FLINT femtosecond laser (1030 nm, 76 MHz, ~150 fs) pumps an OPO to generate a tunable signal (1320–2000 nm); its second harmonic (e.g., 770 nm) is used as the probe. The 1030 nm output is frequency-doubled (SHG) to 515 nm to pump the sample, modulated at 4.73 kHz and temporally delayed via a motorized delay stage. A scanning mirror steers the probe beam across the sample. Pump and probe are combined via a dichroic beamsplitter and focused onto sub-micron spots using a microscope objective. The reflected probe is detected by a Silicon photodiode and demodulated with a lock-in amplifier. Details of the setup and experiment are described in Methods section and in Ref \cite{varghese2023}. BS - Beamsplitter, PD - Photodiode, SHG - Second harmonic generation crystal.}
    \label{fig:ST}
\end{figure}

\begin{figure}[H]
    \centering
    \includegraphics[width=\textwidth]{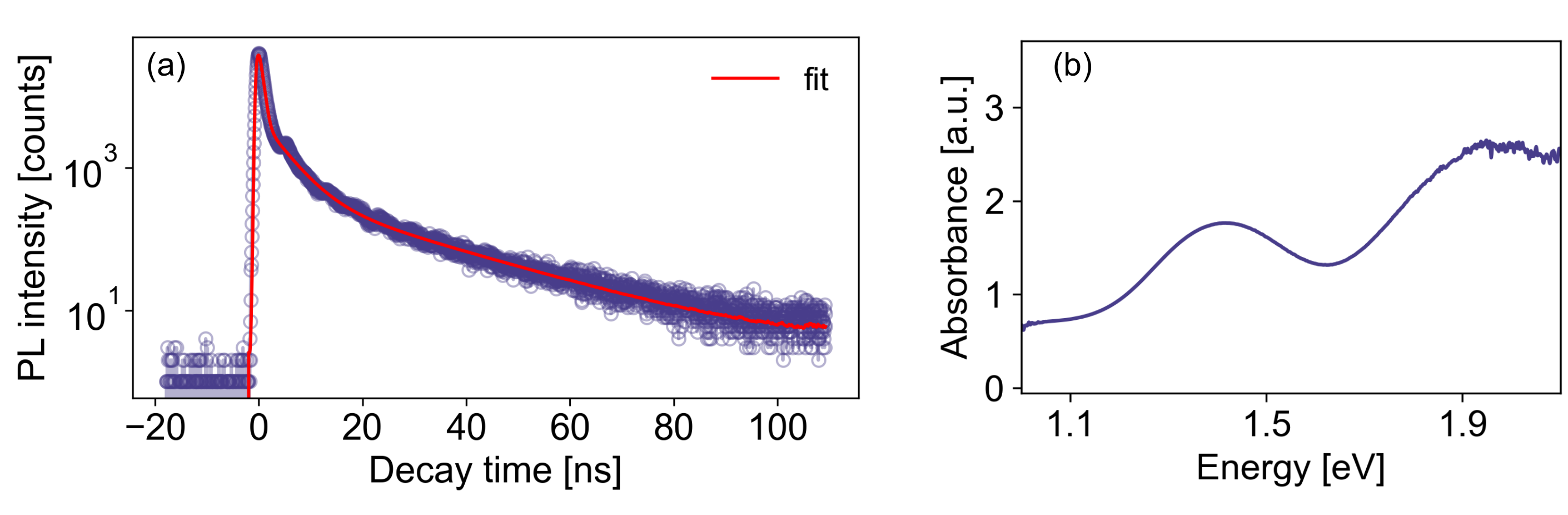}
    \caption{\textbf{Time-resolved photoluminscence and absorption measurements.} 
    \textbf{a)} Time-resolved photoluminescence trace, showing two different decay times, $\tau_{1}$ and $\tau_{2}$. The fast component, $\tau_{1} \sim$ 4.2 ns, represents the radiative recombination time while the slower component $\tau_{2} \sim$ 20.2 ns represents the non-radiative trap-assisted Shockley-Read-Hall recombination. These time scales show that photogenerated charges are relatively long-lived.
    \textbf{b)} Absorbance profile of BiFeO\textsubscript{3}, showing multiple sub-bandgap states with a wide distribution in energy. These defect states mostly originate from off stoichiometry and oxygen vacancies \cite{moubah2012photoluminescence, clark2009energy}.}
    \label{fig:ext3}
\end{figure}

\begin{figure}[H]
    \centering
    \includegraphics[width=\textwidth]{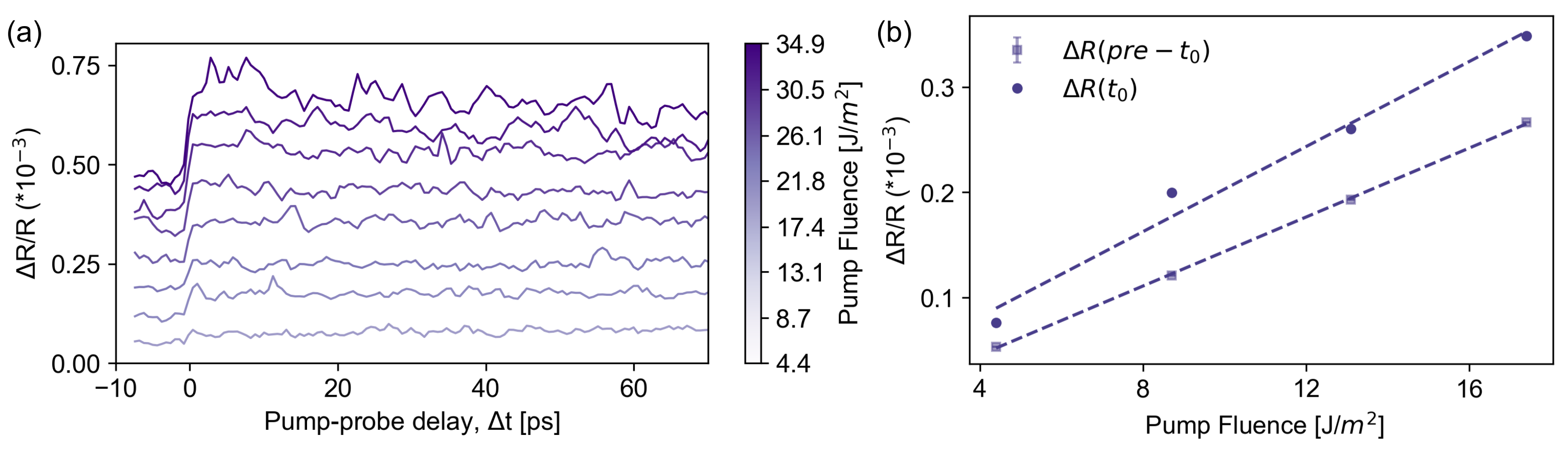}
    \caption{\textbf{Linear regime of the transient reflectivity.} 
    \textbf{a)} By varying the delay time ($\Delta t$) between the arrival of pump and probe pulses, we record the transient change in reflectivity and observe a typical excitation step at $\Delta t_{0}$, which is when the pump and probe pulses overlap temporally, followed by slow relaxation. The non-zero background signal (corresponding to a pump-probe delay time of $\approx$13 ns) originates from the residual excitation in the system from the previous excitation pulse, as is evident from the carrier lifetimes obtained from time-resolved photoluminscence measurements (\ref{fig:ext3}). A larger pump power gives a larger step at time-zero and larger background signal.
    \textbf{b)} Analysis of the pump power-dependent time traces from panel (a), showing linear scaling with pump power, which means that the transient reflectivity represents the carrier density. }
    \label{fig:ext4}
\end{figure}



\begin{figure}[H]
    \centering
    \includegraphics[width=0.5\textwidth]{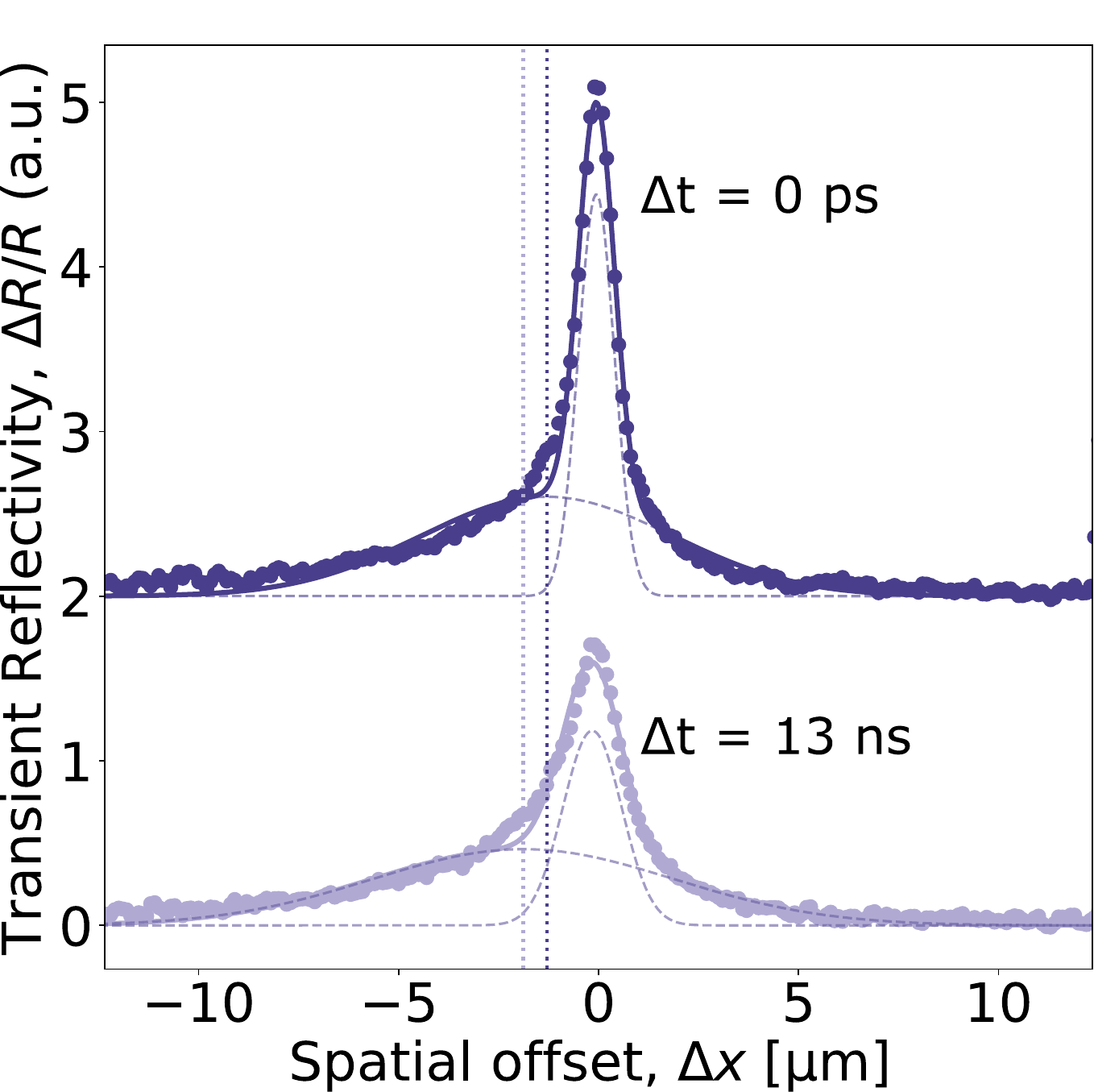}
    \caption{\textbf{Estimation of drift velocity.} Double Gaussian fits to the spatial profiles at $\Delta t = 0$ ps and $\Delta t = 13$ ns. The black dotted lines mark the center of the broader, drifting Gaussian component, revealing a peak displacement of $\sim0.6\,\upmu\mathrm{m}$, corresponding to a drift velocity $v_d \approx 50\,\mathrm{m/s}$. From the broadening of the narrower, diffusing Gaussian component, we obtain a diffusivity of 0.13 cm$^2$/s. This relies on the assumption that all diffusion happens in a time window of 13 ns without accumulation from subsequent pulses in the pulse train.}
    \label{fig:ext6}
\end{figure}

\end{document}